\documentclass[twocolumn,showpacs,preprintnumbers,amsmath,amssymb,prd,floatfix]{revtex4}
\usepackage{graphicx}

\newcommand{\del}[1]{ \partial_{#1} }

\def\al{{\alpha}}
\def\lam{{\lambda}}
\def\gam{{\gamma}}

\def\tilU{{\tilde{U}}}
\def\tU{{\tilde{U}}}
\def\mE{{{\mathcal E}}}

\begin{document}
\title{
Solitonic generation of five-dimensional black ring solution
}
\author{Hideo Iguchi and Takashi Mishima} 
\affiliation{
Laboratory of Physics,~College of Science and Technology,~
Nihon University,\\ Narashinodai,~Funabashi,~Chiba 274-8501,~Japan
}
\date{\today}
\begin{abstract}
Using the solitonic solution-generating technique we rederived the one-rotational
five-dimensional black ring solution found by Emparan and Reall.
The seed solution is not the Minkowski metric, which is the seed
of $S^2$-rotating black ring. The obtained solution has more parameters
than the Emparan and Reall's $S^1$-rotating black ring.
We found the conditions of parameters to reduce the solution
to the $S^1$-rotating black ring.
In addition we examined the relation between the expressions of the metric 
in the prolate-spheroidal coordinates and
in the canonical coordinates.
\end{abstract}
\maketitle


One of the most important recent findings of the higher-dimensional 
General Relativity is a one-rotational black ring solution by Emparan and Reall
\cite{ref1}.
This solution is a vacuum, axially symmetric and asymptotically flat solution
of the five-dimensional General Relativity.
The topology of the event horizon is $S^1 \times S^2$. The black ring rotates
along the direction of the $S^1$. The extension of this solution to 
a two-rotational one has not yet been achieved.

Recently the present authors found a black ring solution with $S^2$ rotation
by using a solitonic solution-generating technique \cite{Mishima:2005id}.
In the analysis we reduced the problem to the four-dimensional one
\cite{{Mazur:1987},{Dereli:1977},{Bruckman:1985yk}}
and applied the formula \cite{ref7} to obtain the metric functions.
The seed solution of this ring is a simple Minkowski spacetime.
Because the effect of rotation cannot compensate for 
the gravitational attractive force, the ring has a kind of strut structure.
Figueras found a C-metric expression of $S^2$-rotating black ring solution
\cite{Figueras:2005zp}.
Tomizawa et al. showed that the same black ring solution
is obtained by using the 
inverse scattering method \cite{Tomizawa:2005wv}.

In this paper we generate the black ring with $S^1$ rotation by 
the solitonic solution-generating technique.
We find that the seed solution is not a Minkowski spacetime.
The obtained solution has more parameters
than Emparan and Reall's black ring. 
It is therefore an extension of the result of Emparan and Reall
and we need some additional conditions to reduce the solution we obtained 
to the black ring solution.
In this analysis we use prolate-spheroidal coordinates. 
The relation between this and the canonical coordinates
considered by Harmark \cite{refHAR} are analyzed.
We also investigate the correspondence between the seed solutions and the 
solitonic ones from the viewpoints of rod structure.
This viewpoint would be helpful to consider seed solutions for 
further new five-dimensional solutions.
We cannot generate two-rotational solutions by the solution-generating 
technique used here. 
However if we use another technique, e.g., inverse scattering method,
for the seed solution used in this analysis or the seed with some corrections, 
the two-rotational black ring solution may be obtained.


At first we briefly explain the procedure to generate axisymmetric solutions
in the five-dimensional general relativity. 
The spacetimes which we considered 
satisfy the following conditions:
(c1) five dimensions, (c2) asymptotically flat spacetimes, 
(c3) the solutions of 
vacuum Einstein equations, (c4) having three commuting Killing vectors 
including time translational invariance and 
(c5) having a single non-zero angular momentum component. 
Under the conditions (c1) -- (c5), 
we can employ the following Weyl-Papapetrou metric form 
(for example, see the treatment in \cite{refHAR}), 
\begin{eqnarray}
ds^2 &=&-e^{2U_0}(dx^0-\omega d\phi)^2+e^{2U_1}\rho^2(d\phi)^2
       +e^{2U_2}(d\psi)^2 \nonumber \\ && 
       +e^{2(\gamma+U_1)}\left(d\rho^2+dz^2\right) ,
       \label{WPmetric}
\end{eqnarray}
where $U_0$, $U_1$, $U_2$, $\omega$ and $\gamma$ are functions of 
$\rho$ and $z$. 
Then we introduce new functions 
$S:=2U_0+U_2$ and $T:=U_2$ so that 
the metric form (1) is rewritten into 
\begin{eqnarray}
ds^2 &=&e^{-T}\left[
       -e^{S}(dx^0-\omega d\phi)^2
       +e^{T+2U_1}\rho^2(d\phi)^2 \right. \nonumber \\&&\hskip 0cm \left.
+e^{2(\gamma+U_1)+T}\left(d\rho^2+dz^2\right) \right]
  +e^{2T}(d\psi)^2.
  \label{MBmetric}
\end{eqnarray}
Using this metric form
the Einstein equations are reduced to the following set of equations, 
\begin{eqnarray*}
&&{\bf\rm (i)}\quad
\nabla^2T\, =\, 0,   \\
&&{\bf\rm (ii)}
\left\{\begin{array}{ll}
& \del{\rho}\gamma_T={\displaystyle
  \frac{3}{4}\,\rho\,
  \left[\,(\del{\rho}T)^2-(\del{z}T)^2\,\right]}\,\ \   \\[3mm]
& \del{z}\gamma_T={\displaystyle 
\frac{3}{2}\,\rho\,
  \left[\,\del{\rho}T\,\del{z}T\,\right],  }
 \end{array}\right.  \\
&&{\bf\rm (iii)}\quad
\nabla^2\mE_S=\frac{2}{\mE_S+{\bar\mE}_S}\,
                    \nabla\mE_S\cdot\nabla\mE_S , \\  
&&{\bf\rm (iv)}
\left\{\begin{array}{ll}
& \del{\rho}\gamma_S={\displaystyle
\frac{\rho}{2(\mE_S+{\bar\mE}_S)}\,
  \left(\,\del{\rho}\mE_S\del{\rho}{\bar\mE}_S
  -\del{z}\mE_S\del{z}{\bar\mE}_S\,
\right)}     \\
& \del{z}\gamma_S={\displaystyle
\frac{\rho}{2(\mE_S+{\bar\mE}_S)}\,
  \left(\,\del{\rho}\mE_S\del{z}{\bar\mE}_S
  +\del{\rho}\mE_S\del{z}{\bar\mE}_S\,
  \right)},  
\end{array}\right.  \\
&&{\bf\rm (v)}\quad
\left( \del{\rho}\Phi,\,\del{z}\Phi \right)
=\rho^{-1}e^{2S}\left( -\del{z}\omega,\,\del{\rho}\omega \right),  \\
&&{\bf\rm (vi)}\quad 
\gamma=\gamma_S+\gamma_T,   \\
&&{\bf\rm (vii)}\quad 
U_1=-\frac{S+T}{2},
\end{eqnarray*}
where $\Phi$ is defined through the equation (v) and the function 
$\mathcal{E_S}$ is defined by 
$
\,\mE_S:=e^{S}+i\,\Phi\,.
$
The most non-trivial task to obtain new metrics is to solve 
the equation (iii) because of its non-linearity. 
To overcome this difficulty 
here we use the method similar to the Neugebauer's 
B\"{a}cklund transformation \cite{{Neugebauer:1980}}
or the Hoenselaers-Kinnersley-Xanthopoulos transformation \cite{Hoenselaers:1979mk}.

To write down the exact form of the metric functions,
we follow the procedure
given by Castejon-Amenedo and Manko \cite{ref7}.
In the five dimensional spacetime we start from the following form of a seed static metric
\begin{eqnarray}
ds^2 &=& e^{-T^{(0)}}\left[
       -e^{S^{(0)}}(dx^0)^2
       +e^{-S^{(0)}}\rho^2(d\phi)^2  \right. \nonumber \\ &&\hskip 0cm \left.
   +e^{2\gamma^{(0)}-S^{(0)}}\left(d\rho^2+dz^2\right) \right]
  +e^{2T^{(0)}}(d\psi)^2.
\nonumber
\end{eqnarray}
For this static seed solution, $e^{S^{(0)}}$, of the Ernst equation (iii), 
a new Ernst potential can be written in the form
\begin{equation}
{\cal E}_S = e^{S^{(0)}}\frac{x(1+ab)+iy(b-a)-(1-ia)(1-ib)}
                         {x(1+ab)+iy(b-a)+(1-ia)(1-ib)},
\nonumber
\end{equation}
where $x$ and $y$ are the prolate spheroidal coordinates:
$
\,\rho=\sigma\sqrt{x^2-1}\sqrt{1-y^2},\ z=\sigma xy\,,
$
with $\sigma>0$. The ranges of these coordinates are 
$1 \le x$ and $-1 \le y \le 1$.
The functions $a$ and $b$ satisfy the following 
simple first-order differential equations 
\begin{eqnarray}
(x-y)\del{x}a&=&
a\left[(xy-1)\del{x}S^{(0)}+(1-y^2)\del{y}S^{(0)}\right], \nonumber \\
(x-y)\del{y}a&=&
a\left[-(x^2-1)\del{x}S^{(0)}+(xy-1)\del{y}S^{(0)}\right], \nonumber \\
(x+y)\del{x}b&=&
-b\left[(xy+1)\del{x}S^{(0)}+(1-y^2)\del{y}S^{(0)}\right] , \nonumber\\
(x+y)\del{y}b&=&
-b\left[-(x^2-1)\del{x}S^{(0)}+(xy+1)\del{y}S^{(0)}\right]. \nonumber \\
\label{eq:ab}
\end{eqnarray}
The metric functions for the five-dimensional metric 
 (\ref{MBmetric}) are obtained
by using the formulas shown by \cite{ref7}, 
\begin{eqnarray}
e^{S}&=&e^{S^{(0)}}\frac{A}{B}   \label{e^S} \\
\omega&=&2\sigma e^{-S^{(0)}}\frac{C}{A}+C_1 \label{omega}     \\
e^{2\gamma}&=&C_2(x^2-1)^{-1}A
                e^{2\gamma'}, \label{e_gamma}
\end{eqnarray}
where $C_1$ and $C_2$ are constants and
$A$, $B$ and $C$ are given by
\begin{eqnarray*}
A&:=&(x^2-1)(1+ab)^2-(1-y^2)(b-a)^2, \\
B&:=&[(x+1)+(x-1)ab]^2+[(1+y)a+(1-y)b]^2, \\
C&:=&(x^2-1)(1+ab)[(1-y)b-(1+y)a]  
\nonumber\\ &&\hskip 0.9cm
+(1-y^2)(b-a)[x+1-(x-1)ab].
\end{eqnarray*}
In addition 
the $\gamma'$ in Eq. (\ref{e_gamma}) is a $\gamma$ function corresponding to the static metric,
\begin{eqnarray}
ds^2 &=& e^{-T^{(0)}}\left[
       -e^{2U^{\mbox{\tiny(BH)}}_0+S^{(0)}}(dx^0)^2
       +e^{-2U^{\mbox{\tiny(BH)}}_0-S^{(0)}}\rho^2(d\phi)^2 \right. 
    \nonumber \\ &&\hskip -0.cm \left.
   +e^{2(\gamma'-U^{\mbox{\tiny(BH)}}_0)-S^{(0)}}\left(d\rho^2+dz^2\right) \right]
  +e^{2T^{(0)}}(d\psi)^2 \label{static_5}
\end{eqnarray}
where ${\displaystyle U_{0}^{\mbox{\tiny(BH)}}=\frac{1}{2}\ln\left( \frac{x-1}{x+1} \right)}$. 
And then the function $T$ is equals to $T^{(0)}$ and $U_1$ is given by 
the Einstein equation (vii).


Using the solution-generating technique described above, we construct
the $S^1$-rotating black ring solution obtained by Emparan and Reall. 
The most important point is to find the seed metric of the 
black ring solution. To do so, it is useful to use
the rod structures which was studied for the higher-dimensional Weyl solutions
by  Emparan and Reall \cite{ref8} and for the nonstatic solutions
by Harmark \cite{refHAR}.
Using this rod structure analysis, 
we found
the seed metric of the $S^1$-rotating black ring solution
by analogy with the relations between
the $S^2$-rotating black ring and its seed metric.
(See \cite{refHAR} for the definition of the rod structure.)

\begin{figure}
  \includegraphics[scale=0.33,angle=0]{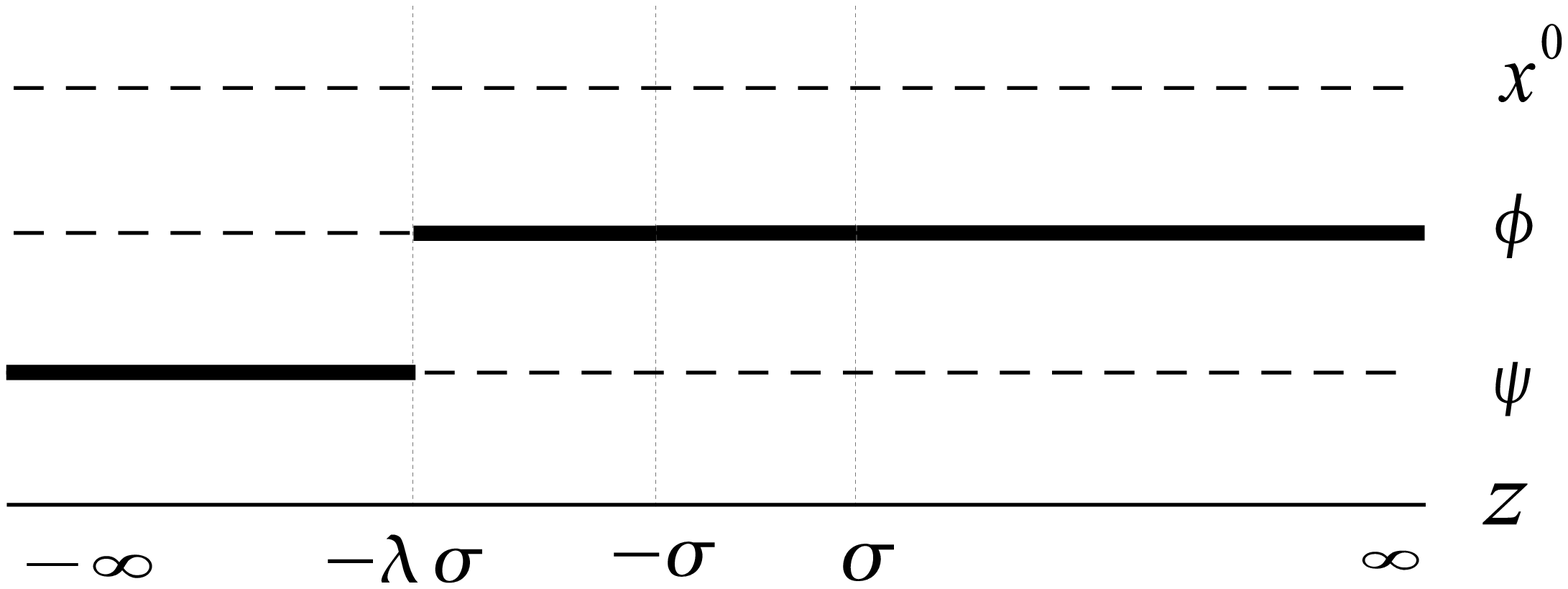}\\
  \includegraphics[scale=0.33,angle=0]{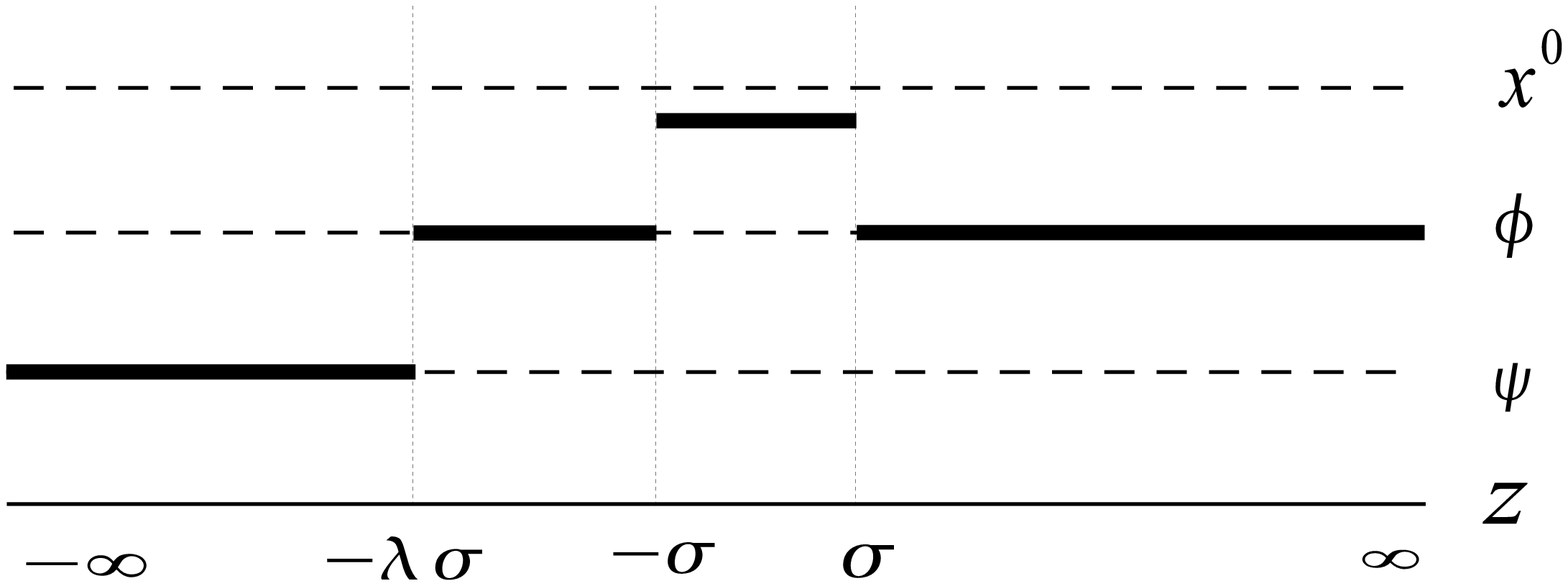}
  \caption{Schematic pictures of rod structures. The upper panel
shows the rod structure of Minkowski spacetime, which is a seed of 
$S^2$-rotating black ring. The lower panel shows the rod structure
of the $S^2$-rotating black ring. The segment $[-\sigma,\sigma]$
of semi-infinite rod in the upper panel is tranformed to 
the finite timelike rod with $\phi$-rotation 
by the solution-generating transformation.
The eigenvector of the finite timelike rod in the lower panel has non-zero 
$\phi$ component. Therefore we put this rod between $x^0$ and $\phi$ axes.
}
 \label{fig:rods_S2}
 \end{figure}

We show the schematic pictures of rod structures of 
the $S^2$-rotating black ring and its seed solution in Fig.
 \ref{fig:rods_S2} \cite{Mishima:2005id}.
Through the solitonic transformation
the segment $[-\sigma,\sigma]$
of semi-infinite spacelike rod which corresponds to the $\phi$-axis 
is changed to the finite timelike rod.
To indicate that the $x^0$ and $\phi$ components of the eigenvector 
are not zero, we put the finite rod between $x^0$ and $\phi$ axes in Fig. \ref{fig:rods_S2}.
In the resulting solution this segment corresponds to
the event horizon with $\phi$-rotation.

The rod structure of $S^1$-rotating black ring was investigated by
Harmark \cite{refHAR}. There are two semi-infinite spacelike rods
in the directions of $\partial/\partial \psi$ and 
$\partial/\partial \phi$. 
Note that these two semi-infinite spacelike rods
assure the asymptotic flatness of the spacetime.
Also there is a finite spacelike rod 
in $\partial/\partial\psi$ direction. A finite timelike rod has finite and
semi-infinite spacelike rods in $\partial/\partial\psi$ direction on each side.
This timelike rod corresponds to an event horizon with $\phi$-rotation.

\begin{figure}
  \includegraphics[scale=0.33,angle=0]{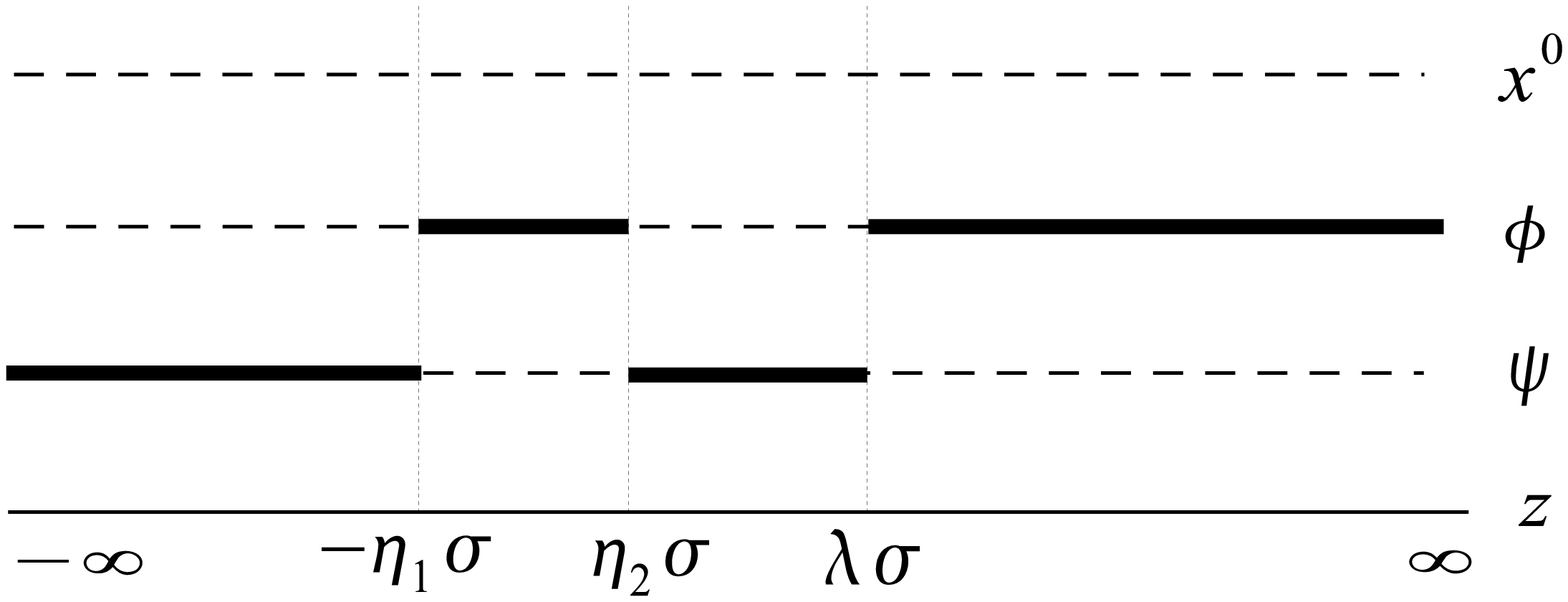}\\
  \includegraphics[scale=0.33,angle=0]{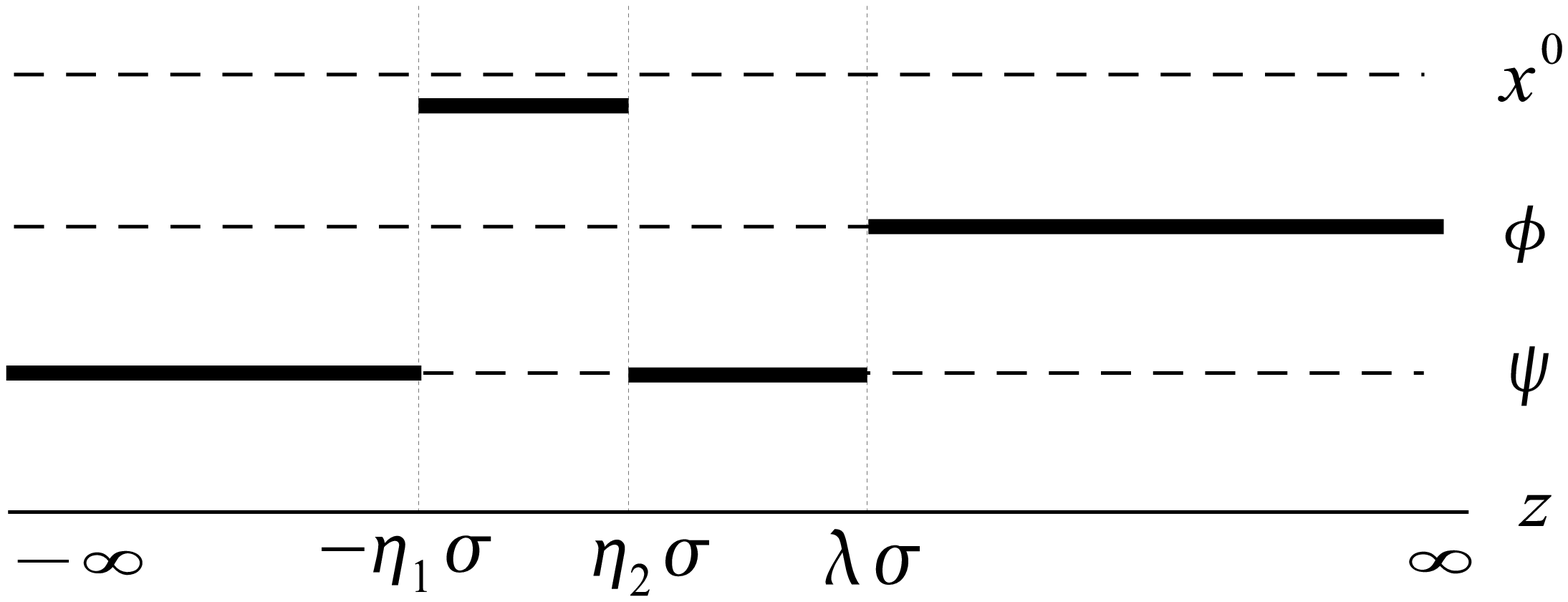}
  \caption{Schematic pictures of rod structures. The upper panel
shows the rod structure of seed metric of 
$S^1$-rotating black ring. The lower panel shows the rod structure
of $S^1$-rotating black ring. The finite spacelike rod
 $[-\eta_1\sigma,\eta_2\sigma]$
in the upper panel is altered to the finite timelike rod by the solution-generating transformation.}
 \label{fig:rods_ER}
 \end{figure}

Now we 
construct the seed solution for the $S^1$-rotating black ring.
For this purpose
we trace back from the rod structure of
$S^1$-rotating black ring
to the seed solution referring the analysis of $S^2$-rotating
black ring.
This can be achieved when we change the
finite timelike rod to the
finite spacelike rod in the $\partial/\partial\phi$ direction.
In Fig. \ref{fig:rods_ER} we show the schematic pictures of
these two rod structures.

The seed metric of $S^1$-rotating black ring is summarized as follows.
The $\psi$-$\psi$ componet of the metric of
black ring and its seed are exactly same as each other.
Also the 0-0 component of the seed metric is $-1$.
The seed functions of 
$S^1$-rotating black ring solution are obtained as
\begin{equation}
S^{(0)}=T^{(0)}= \tilde{U}_{\lambda\sigma} +\tilde{U}_{-\eta_1 \sigma}
                 -\tilde{U}_{\eta_2 \sigma},
      \label{eq:seed} 
\end{equation}
where the function $\tilU_{d}$ is defined as
$\tilU_{d}:=\frac{1}{2}\ln\left[R_{d}+(z-d)\right]$
and $R_d=\sqrt{\rho^2+(z-d)^2}$.
We assume that
the parameters $\lambda$, $\eta_1$ and $\eta_2$ should satisfy
the following inequalities
\begin{eqnarray}
&& 1 \le \lambda,~~-1\le \eta_1\le 1, 
~~~1\le \eta_2\le 1,~~~0\le \eta_1+\eta_2, \nonumber
\end{eqnarray}
to generate the black ring solution
because the timelike rod appears in the region $-\sigma \le z \le \sigma$
after the solitonic transformation.
Under these assumptions, the region of $x=1$ and $-\eta_1<y<\eta_2$
of the solitonic solution corresponds to the event horizon.
Also the region of $x>\lambda$ and $y=1$ is the fixed points of the
$\phi$ rotation.
The regions of $1<x<\lambda$ and $y=1$, $x=1$ and $\eta_2<y<1$,
$x=1$ and $-1<y<-\eta_1$, and $y=-1$ become fixed points of 
$\psi$ rotation in the black ring spacetime.

Substituting the seed function (\ref{eq:seed}) 
into the differential equations (\ref{eq:ab}),
we obtain the solutions of these equations as,  
\begin{equation}
a=\frac{\alpha}{2\sigma^{\frac{1}{2}}}
 \frac{e^{2U_\sigma}+e^{2\tilU_{\lam\sigma}}}{e^{\tilU_{\lam\sigma}}}
 \frac{e^{2U_\sigma}+e^{2\tilU_{-\eta_1\sigma}}}{e^{\tilU_{-\eta_1\sigma}}}
 \frac{e^{\tilU_{\eta_2\sigma}}}{e^{2U_\sigma}+e^{2\tilU_{\eta_2\sigma}}}, \nonumber
\end{equation}
\begin{eqnarray}
&&b={2\sigma^{\frac{1}{2}}}{\beta}
 \frac{e^{\tilU_{\lam\sigma}}}{e^{2U_{-\sigma}}+e^{2\tilU_{\lam\sigma}}}
 \frac{e^{\tilU_{-\eta_1\sigma}}}{e^{2U_{-\sigma}}+e^{2\tilU_{-\eta_1\sigma}}}
 \frac{e^{2U_{-\sigma}}+e^{2\tilU_{\eta_2\sigma}}}{e^{\tilU_{\eta_2\sigma}}},
 \nonumber
\end{eqnarray}
where $\alpha$ and $\beta$ are integration constants and
$U_c:=\frac{1}{2}\ln[R_c-(z-c)]$.

Next we reduce the explicit expression for the $\gamma'$.
Read out the functions $S'$ and $T'$ from Eq. (\ref{static_5}) as
\begin{eqnarray}
S'&=&2\,U^{(BH)}_0+S^{(0)} \nonumber \\
  &=& 2(\tilU_\sigma-\tilU_{-\sigma})
   +\tilU_{\lam\sigma}+\tilU_{-\eta_1\sigma}-\tilU_{\eta_2\sigma} \nonumber\\
T'&=&T^{(0)}=\tilU_{\lam\sigma}+\tilU_{-\eta_1\sigma}-\tilU_{\eta_2\sigma},
\nonumber
\end{eqnarray}
and substitute them into
\begin{equation}
\del{\rho}\gam'=
\frac{1}{4}\rho\left[(\del{\rho}S')^2-(\del{z}S')^2\right]
 +\frac{3}{4}\rho\left[(\del{\rho}T')^2-(\del{z}T')^2\right],
\nonumber
\end{equation}
\begin{equation}
\del{z}\gam'=
\frac{1}{2}\rho\left[\del{\rho}S'\del{z}S'\right]
 +\frac{3}{2}\rho\left[\del{\rho}T'\del{z}T'\right],
\nonumber
\end{equation}
so we can confirm that $\gamma'$ is divided 
as
\begin{eqnarray}
\gam' &=& \gam'_{\sigma,\sigma}+\gam'_{-\sigma,-\sigma}
+\gam'_{\lambda\sigma,\lambda\sigma}+\gam'_{-\eta_1\sigma,-\eta_1\sigma}
+\gam'_{\eta_2\sigma,\eta_2\sigma} \nonumber \\
&& 
 -2\gam'_{\sigma,-\sigma}
+\gam'_{\sigma,\lambda\sigma}+\gam'_{\sigma,-\eta_1\sigma}
-\gam'_{\sigma,\eta_2\sigma}
\nonumber \\ &&
-\gam'_{-\sigma,\lambda\sigma}
-\gam'_{-\sigma,-\eta_1\sigma}
+\gam'_{-\sigma,\eta_2\sigma}
+2\gam'_{\lam\sigma,-\eta_1\sigma}
\nonumber \\ &&
-2\gam'_{\lam\sigma,\eta_2\sigma}
-2\gam'_{-\eta_1\sigma,\eta_2\sigma} ,
\nonumber
\end{eqnarray}
where $\gamma'_{cd}$ satisfies the following equations
\begin{eqnarray}
\del{\rho}\gam'_{cd}
    &=&\rho\left[\del{\rho}\tU_{c}\del{\rho}\tU_{d}
            -\del{z}\tU_{c}\del{z}\tU_{d}\right],  \label{eq:drho_gm}\\
\del{z}\gam'_{cd}
    &=&\rho\left[\del{\rho}\tU_{c}\del{z}\tU_{d}
           +\del{\rho}\tU_{d}\del{z}\tU_{c}\right]. \label{eq:dz_gm}
\end{eqnarray}
These equations (\ref{eq:drho_gm}) and (\ref{eq:dz_gm}) have the solution, 
\begin{equation}
\gam'_{cd}=\frac{1}{2}\tU_{c}+\frac{1}{2}\tU_{d}-\frac{1}{4}\ln Y_{cd}, 
\nonumber 
\end{equation}
where $Y_{cd}:=R_cR_d+(z-c)(z-d)+\rho^2$. 

Now the functions which is needed to express the full metric
are completely obtained.
The full metric is expressed as 
\begin{eqnarray} 
ds^2 &=&-\frac{A}{B}\left[dx^0-\left(2\sigma e^{-S^{(0)}}
         \frac{C}{A}+C_1\right) d\phi\right]^2  \nonumber  \\ &&
+\frac{B}{A}e^{-2S^{(0)}}\rho^2(d\phi)^2 
+e^{2S^{(0)}}(d\psi)^2
                  \nonumber \\ && 
+C_2 \sigma^2 \frac{x^2 - y^2}{x^2 -1}B e^{2(\gam'-S^{(0)})}
\left(\frac{dx^2}{x^2-1}+\frac{dy^2}{1-y^2}\right).
\nonumber \\ &&
\label{eq:metric_ER}
\end{eqnarray}
In the following
the constants $C_1$ and $C_2$ are fixed as 
\[
C_1=\frac{\,\,2\sigma^{1/2}\,\al\,\,}{1+\al\beta},\ \ \ 
C_2=\frac{1}{\sqrt{2}(1+\al\beta)^2},
\]
to assure that the spacetime asymptotes to a five-dimensional 
Minkowski spacetime globally.
We can confirm this by taking the asymptotic
limit, $x \rightarrow \infty$, of the metric.

The above solution is an extension of $S^1$-rotating black ring solution
because it contains singular cases.
In general the spacetime described by the metric (\ref{eq:metric_ER})
includes some harmful regions, for example,  the region
where closed timelike curves exist.
In fact, the metric component
$g_{\phi\phi}$ becomes negative around $(x,y)=(1,1)$ and $(1,-1)$.
These singular behaviors are cured by setting the parameters $\alpha$ and 
$\beta$ as
\begin{eqnarray}
\alpha =  \sqrt{\frac{2(1-\eta_2)}{(\lambda -1)(1+\eta_1)}}, \hspace{0.2cm}
 \beta =  \sqrt{\frac{(\lambda +1)(1-\eta_1)}{2(1+\eta_2)}} . 
 \label{eq:beta} 
\end{eqnarray}

The asymptotic form of $\mE_{S}$ near the infinity $\tilde{r}=\infty$ becomes 
\begin{eqnarray}
\mE_{S}&=&\tilde{r}\cos\theta\,
\left[\,1\,-\,\frac{\sigma}{\tilde{r}^2}\,\frac{P(\al,\beta,\lam)}
                                       {(1+\alpha\beta)^2}
     \,+\cdots\right] \nonumber  \\ &&  \hskip -0cm   
     +2\,i\,\sigma^{1/2}\,\left[\,\frac{\alpha}{1+\alpha\beta}
      \,-\,\frac{2\sigma\cos^2\theta}{\tilde{r}^2}\,\frac{Q(\al,\beta,\lam)}
                                                 {(1+\alpha\beta)^3}
    \,\,+\cdots\,\right],
\nonumber 
\end{eqnarray}
where we introduced the new coordinates $\tilde{r}$ and $\theta$ through
the relations
\begin{equation}
 x=\frac{\tilde{r}^2}{2\sigma}+\lambda-\eta_1-\eta_2, \ y=\cos 2\theta,
\nonumber
\end{equation}
and 
\begin{eqnarray}
 P(\al,\beta,\lam)
   &=&  4(1 + \alpha^2 - \alpha^2 \beta^2)     \nonumber  \\
Q(\al,\beta,\lam)
   &=& \alpha(2\alpha^2-\eta_1-\eta_2+\lam+3)-2\alpha^2\beta^3  \nonumber\\
    && 
  -\beta\left[2(2\alpha\beta+1)(\alpha^2+1) \right. \nonumber\\
    && \left.
  +(\eta_1+\eta_2-\lam-1)\al^2(\al\beta+2)\right].  \nonumber
\end{eqnarray}
 From the asymptotic behavior of the Ernst potential, 
 we can compute the mass
parameter $m^2$ and rotational parameter $m^2a_0$ as
\begin{eqnarray}
&& m^2 = \sigma \frac{P(\al,\beta,\lam)}{(1+\alpha\beta)^2}, \hspace{0.2cm}
m^2a_0 = 4\sigma^{3/2}\frac{Q(\al,\beta,\lam)}{(1+\alpha\beta)^3}. \nonumber 
\end{eqnarray}
For the black ring solution we obtain
\begin{eqnarray}
 m^2 &=&
 \frac{8 \sigma (\eta_1 + \eta_2)(\lambda-\eta_2)}
            {(\lambda-1)(1+\eta_1)(1+\eta_2)(1+\alpha\beta)^2}, \label{eq:m2_S1}\\
 m^2a_0 &=&
  m^2\frac{\sqrt{\sigma}(\alpha(1+\lambda\eta_1)-2\beta\eta_2)}{1+\alpha\beta},
\label{eq:m2a0_S1}
\end{eqnarray}
where we use the Eq. 
(\ref{eq:beta}).
 From Eqs. (\ref{eq:m2_S1}) and (\ref{eq:m2a0_S1}) 
we can obtain the usefull relation
\begin{equation}
 \frac{a_0^2}{m^2} = \frac{\left((1+\lambda\eta_1)\sqrt{1-\eta_2^2}-
                         \eta_2\sqrt{(\lambda^2-1)(1-\eta_1^2)}\right)^2}
 {4(\lambda-\eta_2)(\eta_1+\eta_2)}.
\label{eq:a0m}
\end{equation}
When $\eta_1=\eta_2=1$
the  black ring becomes static because $\alpha=\beta=0$ from
Eq. 
(\ref{eq:beta}).
The one-rotational black hole limit \cite{Myers:1986un}
of the black ring
is realized when we set 
\begin{eqnarray}
 \lambda=1+\epsilon, ~~~ \eta_2=1-k\epsilon,
 \nonumber
\end{eqnarray}
where $k>0$ is constant,
and then, take the limit $\epsilon \rightarrow 0$.

The periods of $\phi$ and $\psi$ are defined as
\begin{eqnarray}
&& \Delta \phi = 2 \pi \lim_{\rho \rightarrow 0} \sqrt{\frac{\rho^2 g_{\rho\rho}}{g_{\phi\phi}}} 
 ~~~\mbox{and}~~~
 \Delta \psi = 2 \pi \lim_{\rho \rightarrow 0} \sqrt{\frac{\rho^2 g_{\rho\rho}}{g_{\psi\psi}}} \nonumber 
\end{eqnarray}
to avoid a conical singularity.
We see that the period of $\phi$ is
 $\Delta \phi = 2 \pi$ 
and the period of $\psi$ is
 $\Delta \psi = 2 \pi$ 
outside the ring
and
\begin{equation}
 \Delta \psi = 2\pi \frac{
           \left(\frac{\lam - \eta_2}{\lam + \eta_1}\right)
           \left(1 +  \sqrt{
                       \frac{(\lam-1)(1-\eta_1)(1-\eta_2)}
                            {(\lam+1)(1+\eta_1)(1+\eta_2)}}\right)}
           {\sqrt{\frac{\lam - 1}{\lam +1}}+\sqrt{
                       \frac{(1-\eta_1)(1-\eta_2)}
                            {(1+\eta_1)(1+\eta_2)}}}, \nonumber
\end{equation}
inside the ring. In general there is a conical singularity inside or 
outside the ring. This conical singularity is cured by setting the 
parameters $\lambda$, $\eta_1$ and $\eta_2$ to satisfy the relation
\begin{eqnarray}
       &&   \left(\frac{\lam - \eta_2}{\lam + \eta_1}\right)
           \left(1 +  \sqrt{
                       \frac{(\lam-1)(1-\eta_1)(1-\eta_2)}
                            {(\lam+1)(1+\eta_1)(1+\eta_2)}}\right)
      \nonumber \\
    && =\sqrt{\frac{\lam - 1}{\lam +1}}+\sqrt{
                       \frac{(1-\eta_1)(1-\eta_2)}
                            {(1+\eta_1)(1+\eta_2)}}.
   \label{eq:balance}
\end{eqnarray}

Finaly we 
consider the coordinates transformation between
the prolate-spheroidal coordinates used here and 
the canonical coordinates analyzed by Harmark \cite{refHAR}.
See \cite{refHAR} for the notaiton and the exact expression of 
the metric in the canonical coordinates.
We however use $\tilde{\rho}$ and $\tilde{z}$ for $\rho$ and $z$ of \cite{refHAR}.

Comparing the functional forms of $\psi$-$\psi$ components, we obtain
the relations between these two coordinates.
These two coordinates can be transformed into each other
through the relation
\begin{eqnarray}
 \tilde{\rho}&=&\rho, \nonumber \\
 \tilde{z} &=& z +\frac{\eta_1-\eta_2}{2}\sigma 
       =\sigma \left(x y+\frac{\eta_1-\eta_2}{2}\right). \nonumber
\end{eqnarray}
In addition,
the parameters should satisfy the following relations, 
\begin{eqnarray}
 \kappa^2 &=& \sigma\left(\lambda + \frac{\eta_1-\eta_2}{2}\right),
  \label{eq:k2}\\
 c &=& \frac{\eta_1+\eta_2}{2\lambda+\eta_1-\eta_2}, \label{eq:2}\\
 b &=& \frac{(\lambda+1+(\lambda-1)\alpha\beta)^2-(\lambda^2-1)(1+\alpha\beta)^2}
            {(\lambda+1+(\lambda-1)\alpha\beta)^2+(\lambda^2-1)(1+\alpha\beta)^2},
  \label{eq:b}
\end{eqnarray}
to assure the equivalence of these two expressions.
Here the parameters $\alpha$ and $\beta$ satisfy
the conditions 
(\ref{eq:beta}).
Also we have to rescale the $\rho$-$z$ part of the metric as
\begin{equation}
 e^{2\nu}= \frac{1-b}{(1-c)^2} e^{2(\gamma + U_1)}. \nonumber
\end{equation}
%
Note that $b \ge c$ when $\lambda \ge 1$, $\eta_1 \le 1$ and 
$\eta_2 \le 1$. From the static black ring condition $b=c$ \cite{refHAR},
we can derive the following relation
\begin{equation}
 \frac{1-\eta_2}{1+\eta_2}=\left(\frac{1-\eta_1}{1+\eta_1}\right)
                           \left(\frac{\lambda-1}{\lambda+1}\right),
\nonumber
\end{equation} 
which holds when $\eta_1=\eta_2=1$. Indeed the black ring is static
in this case because $\alpha=\beta=0$ from Eq.
(\ref{eq:beta}).
The black ring solution 
becomes one-rotational black hole when we take the 
limits $b,c \rightarrow 1$ \cite{refHAR}.
These limits surely correspond to the limits 
$\lambda, \eta_2 \rightarrow 1$.

There are six parameters $(\lambda,\eta_1,\eta_2,\sigma,\alpha,\beta)$ 
in the metric (\ref{eq:metric_ER}) with two relations 
(\ref{eq:beta}). While there are three parameters $(b,c,\kappa)$
in the canonical coordinates. 
The parameter $\sigma$ 
appears only in the relation of $\kappa^2$, Eq. (\ref{eq:k2}),
and contributes to the scaling of the coordinates.
Thus we can freely fix one of parameters $(\lambda,\eta_1,\eta_2)$.
Here we set $\eta_1=1$ because the relations obtained above 
become simple. In this case 
we can inversely solve Eqs. (\ref{eq:k2}) - (\ref{eq:b}).
The results are
\begin{eqnarray}
&& \eta_1 = 1, ~
 \eta_2 = \frac{2c+cb-b}{(1+c)b} ,~ 
 \lambda = \frac{1}{b},~ 
 \sigma = \frac{1+c}{1+b} b\kappa^2. \nonumber
\end{eqnarray}
When the black ring does not have a conical singularity,
these relations become
\begin{eqnarray}
&& \eta_1 = 1 ,~
 \eta_2 = c,~
 \lambda 
         = \frac{1+c^2}{2c} ,~ 
 \sigma = \frac{2c}{1+c} \kappa^2, \nonumber
\end{eqnarray}
and the parameter $\eta_2$ lies in the range $0<\eta_2\le1$. 

Next we consider the relations between physical variables.
The ADM mass and angular momentum of the black ring were derived by 
Emparan and Reall \cite{ref1} as
\begin{equation}
 M = \frac{3\pi\kappa^2 b}{2(1-c)},~~~
 J_1=\sqrt{2}\pi\kappa^3\frac{\sqrt{b(b-c)(1+b)}}{(1-c)^2}. \nonumber
\end{equation}
These two variables are related to the mass and rotational parameters as
\begin{equation}
 m^2 =\frac{8(1-c)^2}{3\pi(1-b)}M, ~~~
 m^2a_0=\frac{4}{\pi}\frac{(1-c)^3}{(1-b)^{\frac{3}{2}}}J_1.
\nonumber
\end{equation}
Using the balanced black ring conditions (\ref{eq:balance}) with $\eta_1=1$,
 the right hand side of Eq. (\ref{eq:a0m})
can be reduced to the following form,
\begin{equation}
 \frac{a_0^2}{m^2}=\frac{(1+\eta_2)^3}{8\eta_2}
                  =\frac{27\pi}{32}\frac{J_1^2}{M^3}.
\nonumber
\end{equation}
This agrees with the previous results \cite{ref1}.


In this paper we rederived the $S^1$-rotating black ring solution
by the solitonic slution-generating technique.
Using the rod structure analysis
we found the seed solution of the black ring on the analogy
of the relation between the $S^2$-rotating black ring and its seed
solution. The relations between the seed and obtained solitonic solutions
can be easily understood through the analysis of their rod structures.
Thus the rod structure analysis is expected to be a useful guide
to construct seed solutions for new solutions.
In addition 
we obtained the relations between the prolate-spheroidal coordinates
and the canonical coordinates.
This means that using
the coordinates transformation between 
canonical and C-metric coordinates obtained by Harmark \cite{refHAR},
the prolate-spheroidal coordinates also
can be transformed into the C-metric coordinates.

As is the $S^2$-rotating black ring \cite{Tomizawa:2005wv}, 
the $S^1$-rotating black ring solution
would be generated from the seed solution obtained here 
by the inverse scattering method.
Also it might be expected that the two-rotational black ring is obtained 
from the seed solution or the seed with some deformations 
by this method.

This work is partially supported by Grant-in-Aid for Young Scientists (B)
(No. 17740152) from Japanese Ministry of Education, Science,
Sports, and Culture
and by Nihon University Individual Research Grant for
2005.



%
\end{document}